\documentclass{svproc}
\usepackage{url}
\usepackage{graphicx}
\usepackage{epstopdf}

\begin{document}
\mainmatter              
\title{Production of muons from heavy-flavour hadron decays in heavy-ion collisions with ALICE at the LHC}
\author{Bharati~Naik for the ALICE Collaboration}
\titlerunning{HFM production}
\institute{University of the Witwatersrand, Johannesburg, South Africa\\
iThemba LAB, Cape Town, South Africa\\
\email{bharati.naik@cern.ch}
}
\maketitle              
\begin{abstract}
Measurements of the production of muons from heavy-flavour hadron decays at forward rapidity ($2.5 < y < 4.0$) in Pb--Pb collisions at $\sqrt{s_{\rm NN}} = 2.76$ and 5.02 TeV with the ALICE detector are presented along with the $R_{\rm AA}$ measurements in different centrality intervals as a function of transverse momentum, $p_{\rm T}$. Results of the measured $R_{\rm AA}$ at both energies are shown and are compared to the different model predictions.
\end{abstract}
\section{Introduction}
\label{intro}
Heavy quarks (charm and beauty) are produced in the early stages of high-energy heavy-ion collisions via hard parton scattering processes due to their heavy masses. As a result, they experience the entire evolution of the system formed in such collisions. During their propagation through the quantum chromodynamics (QCD) medium, they lose energy via collisional and radiative processes. Therefore, they are effective probes for investigating the properties of the hot and dense QCD medium. The main observable to study the medium effects is the nuclear modification factor $R_{\rm AA}$ \cite{HFMPbPb}, which is expressed as:
\begin{equation}
\label{eq:RAA}
R_{\rm AA}(p_{\rm T}, y) = \frac{1}{\langle T_{\rm AA}\rangle}\times \frac{{\rm d^2}N_{\rm AA}/{\rm d}p_{\rm T}dy}{{\rm d^2}\sigma_{\rm pp}/{\rm d}p_{\rm T}dy} 
\end{equation}
where, ${\rm d^2}N_{\rm AA}/{\rm d}p_{\rm T}dy$ is the $p_{\rm T}$ and $y$-differential particle yields in nucleon\textendash nucleon (AA) collisions and ${\rm d^2}\sigma_{\rm pp}/{\rm d}p_{\rm T}dy$ is the corresponding production cross section in proton--proton (pp) collisions. The nuclear overlap function, $\langle T_{\rm AA}\rangle$, is calculated using the average number of nucleon-nucleon collisions $\langle N_{\rm coll}\rangle$ and the inelastic nucleon-nucleon cross section. The measurement of $R_{\rm AA}$ can provide information about any possible modification of heavy-quark hadronization and in-medium parton energy loss mechanisms.
\section{Analysis method}
\label{analysis}
The main detectors used for this analysis are the forward muon spectrometer, located at a pseudorapidity $-4 < \eta < -2.5$ , to detect the muons, the Silicon Pixel Detector (SPD) for the identification of collision vertex, and the two V0 detectors ($2.8<\eta < 5.1$, $-3.7 < \eta < -1.7$) for triggering minimum bias (MB) events as well as for centrality determination. The various sub-systems of the ALICE apparatus and their performance are described in detail in \cite{RefJ,RefJ1}. The data sets used for the Pb--Pb analysis at $\sqrt{s_{\rm NN}} = 5.02$ TeV ($\sqrt{s_{\rm NN}} = 2.76$ TeV) were collected for muon low-$p_{\rm T}$ trigger thresholds (MSL) of 1.0 (0.5) GeV/$c$ and muon high-$p_{\rm T}$ trigger thresholds (MSH) of 4.2 GeV/$c$. The corresponding integrated luminosities $L\textsubscript{int}$ were about 21.9 (224.8) $\mu$b\textsuperscript{-1} and 4.0 (71.0) $\mu$b\textsuperscript{-1} for MSL (MSH) trigger events after the event selections at $\sqrt{s_{\rm NN}} = 5.02$ and 2.76 TeV, respectively.\\
Muon tracks are reconstructed within  $-4 < \eta < -2.5$, which is the acceptance region of the muon spectrometer. After applying all the muon track selection criteria discussed in \cite{HFMPbPb}, the inclusive muon yields in different centrality classes (0--10$\%$, 20--40$\%$, 60--80$\%$) at  $\sqrt{s_{\rm NN}} = 5.02$ TeV and in the 0-10$\%$ centrality class at $\sqrt{s_{\rm NN}} = 2.76$ TeV are obtained. The inclusive muon yields in Pb--Pb collisions are corrected for detector acceptance and efficiencies and further normalized to the number of equivalent MB events. The inclusive muon yields also contain the background of muons from primary and secondary pion and kaon decays, W-boson decays, Z/$\gamma^*$ decays, and $J/\psi$ decays. After subtracting all background contributions from the inclusive muon yield, the yield of muons from heavy-flavour hadron decays is obtained. The $p_{\rm T}$-differential $R_{\rm AA}$ of muons from the heavy-flavour hadron is calculated using Eq.~\ref{eq:RAA}. The differential production cross section in pp collisions in the denominator of Eq.~\ref{eq:RAA} are taken from the same kinematic region and center-of-mass energy as in Pb--Pb collisions \cite{HFMpp}.


\section{Results}
\label{result}
\begin{figure}[htb!]
\centering
\includegraphics[width=0.6\textwidth,clip]{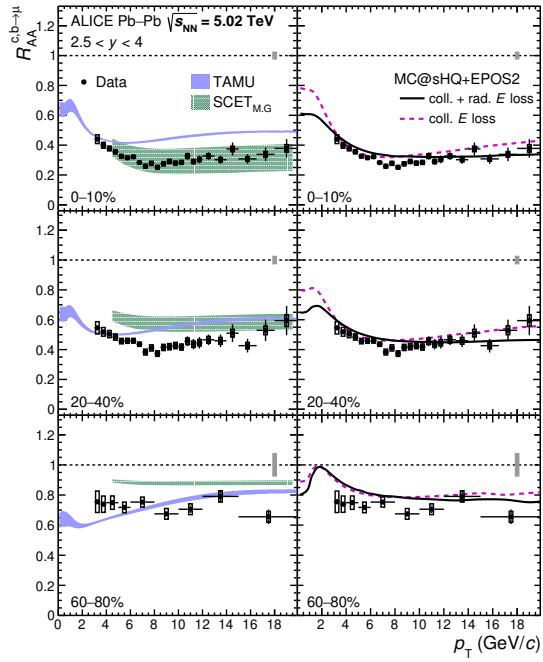}
\caption{$p_{\rm T}$-differential $R_{\rm AA}$ of muons from heavy-flavour hadron decays in central ($0-10\%$), semi-central ($20-40\%$) and peripheral ($60-80\%$) Pb--Pb collisions at $\sqrt{s_{\rm NN}} = 5.02$ TeV \cite{HFMPbPb}.}
\label{fig-RAA502}       
\end{figure}

The $p_{\rm T}$-differential $R_{\rm AA}$ of muons from heavy-flavour hadron in Pb--Pb collisions at $\sqrt{s_{\rm NN}} = 5.02$ TeV over a wide $p_{\rm T}$ range (3 to 20 GeV/$c$) is shown in Fig.~\ref{fig-RAA502}.
The $R_{\rm AA}$ increases from central to peripheral collisions. The suppression is larger in the region $6 < p_{\rm T} < 10$ GeV/$c$ for central and semi-central collisions, while there is no significant $p_{\rm T}$ dependence found in peripheral collisions. The strong suppression observed in Pb--Pb collisions is due to final-state interactions of charm and beauty quarks in the QCD medium. The $R_{\rm AA}$ is compared with different model predictions such as TAMU \cite{model1}, SCET \cite{model2}, and MC@sHQ+EPOS2 \cite{model3,model4}. The TAMU model, which considers only the elastic collisions, underestimates the suppression in central and semi-central collisions.
In contrast, the pQCD-based SCET model considers the medium-induced gluon radiation and describes the data in central collisions. The MC@sHQ+EPOS2 model with and without radiative energy loss describes the data within uncertainties for all centralities over the entire $p_{\rm T}$ interval. 
\begin{figure}[htb!]
\centering
\includegraphics[width=0.5\textwidth,clip]{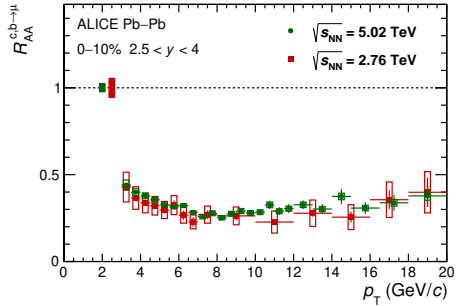}
\caption{Comparison of $p_{\rm T}$ differential $R_{\rm AA}$ in central Pb--Pb collisions at $\sqrt{s_{\rm NN}} = 5.02$ TeV and 2.76 TeV \cite{HFMPbPb}.}
\label{fig-RAAComp}       
\end{figure}
Figure~\ref{fig-RAAComp} shows the comparison between the measured $R_{\rm AA}$ in central Pb--Pb collisions at $\sqrt{s_{\rm NN}} = 5.02$ TeV and 2.76 TeV. The similar suppression in Fig.~\ref{fig-RAAComp} for both energies may result from the interplay between two effects as discussed in \cite{theory}. One is the flattening of the $p_{\rm T}$ spectra of charm and beauty quarks with increasing collision energy which would decrease the heavy-quark suppression (increase $R_{\rm AA}$) by about $5\%$ if the medium temperature remains unchanged. Another effect is the medium temperature estimated to be higher by about 7$\%$ in Pb--Pb collisions at $\sqrt{s_{\rm NN}} = 5.02$ TeV than at 2.76 TeV, which would increase the suppression (decrease $R_{\rm AA}$) by 10$\%$ ($5\%$) for charm (beauty) quarks. 
 
\begin{figure}[htb!]
\centering
\includegraphics[width=0.6\textwidth,clip]{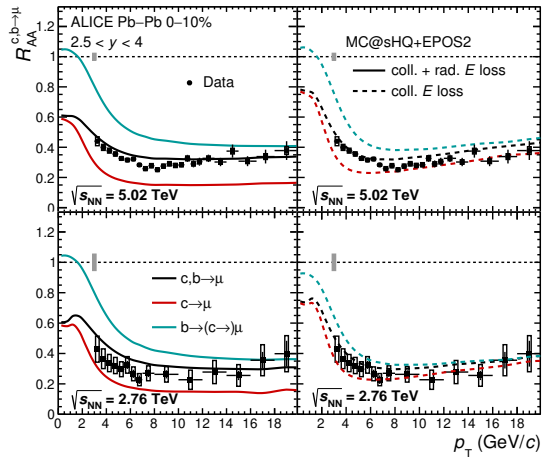}
\caption{Comparison of $p_{\rm T}$ differential $R_{\rm AA}$ with  MC@sHQ+EPOS2 model in central Pb--Pb collisions at $\sqrt{s_{\rm NN}} = 5.02$ TeV (top) and 2.76 TeV (bottom) \cite{HFMPbPb}.}
\label{fig-RAAmodel}       
\end{figure}
The $R_{\rm AA}$ measured in central (0--10$\%$) Pb--Pb collisions at $\sqrt{s_{\rm NN}} = 5.02$ TeV and 2.76 TeV is compared in Fig.~\ref{fig-RAAmodel} with predictions from MC@sHQ+EPOS2 for muons from charm- and beauty-hadron decays, separately, and for muons from the combination of the two. 
The model predictions are closer to the data at high $p_{\rm T}$ for muons from beauty-hadron decays than for muons from charm-hadron decays when considering both elastic and radiative energy loss. However, when the model considers only collisional energy loss mechanisms, the difference between the suppression of muons from beauty and charm-hadron decay is less pronounced. It is worth mentioning that the model is characterized by a large running coupling constant $\alpha_{\rm s}$ and a reduced Debye mass in the elastic heavy quark scattering generating the radiation. As a result, the model describing both collisional and radiative energy loss overestimates the data in high $p_{\rm T}$, as the radiative energy loss neglects the finite path-length effects due to the gluon formation outside the QCD medium. Due to the dead-cone effect, this effect is expected to be more pronounced for charm quarks than beauty quarks \cite{HFMPbPb}.

%
%

\end{document}